\documentclass[preprint,prd,aps,showpacs,showkeys,nofootinbib]{revtex4-1}
\usepackage{amssymb}
\usepackage{amsmath}
\usepackage{graphicx}
\usepackage{subfigure}
\usepackage{float}
\usepackage{mathrsfs}
\usepackage{amsmath}
\begin{document}

\title{The LISA-Taiji network}

\author{Wen-Hong Ruan}
\email{ruanwenhong@itp.ac.cn}

\author{Chang Liu}
\email{liuchang@itp.ac.cn}

\author{Zong-Kuan Guo}
\email{guozk@itp.ac.cn}

\author{Yue-Liang Wu}
\email{ylwu@itp.ac.cn}

\author{Rong-Gen Cai}
\email{cairg@itp.ac.cn}

\affiliation{CAS Key Laboratory of Theoretical Physics,
Institute of Theoretical Physics, Chinese Academy of Sciences, P.O. Box 2735, Beijing 100190, China}

\affiliation{School of Physical Sciences,
University of Chinese Academy of Sciences, No. 19A Yuquan Road, Beijing 100049, China}

\begin{abstract}
Both LISA and Taiji, planned space-based gravitational-wave detectors in orbit around the Sun, are expected to launch in 2030-2035. Assuming a one-year overlap, we explore a potential LISA-Taiji network to fast and accurately localize the gravitational-wave sources.
\end{abstract}


\maketitle

\begin{figure}[!h]
  \centering
  \includegraphics[width=5in]{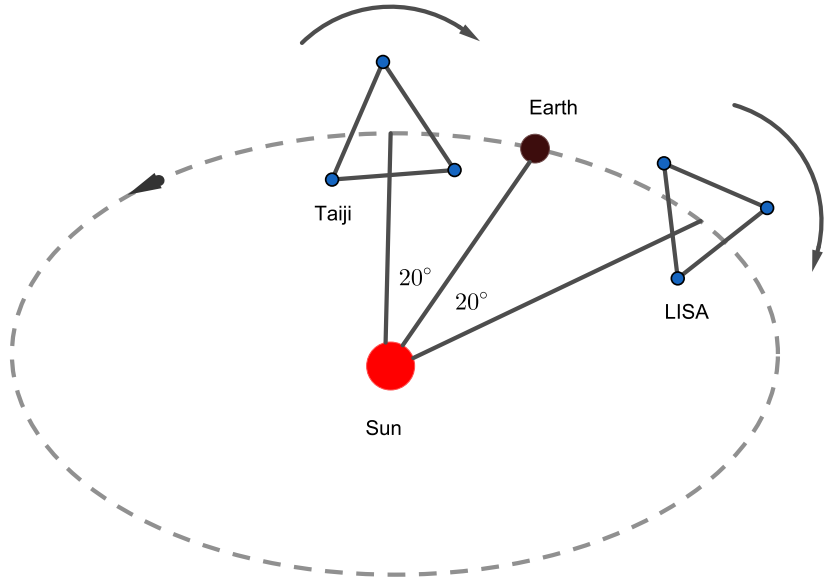}
  \caption{Configuration of the LISA-Taiji network.
  LISA consists of a triangle of three spacecraft with a separation distance of 2.5 million kilometers
  in a heliocentric orbit behind the Earth by about $20^{\circ}$
  while Taiji has a separation distance of 3 million kilometers in a heliocentric orbit ahead of the Earth by about $20^{\circ}$.
  Due to a distance of about 0.7 AU between the two constellations,
  the LISA-Taiji network is expected to significantly improve the sky localization of coalescing massive black hole binaries.}
  \label{fig:configuration}
\end{figure}

Laser Interferometer Space Antenna (LISA), a space-based gravitational-wave observatory,
was proposed in 1990s to detect gravitational waves with a frequency band from $10^{-4}$ Hz to $10^{-1}$ Hz~\cite{Audley:2017drz}.
LISA consists of a triangle of three identical spacecraft
with a separation distance of 2.5 million kilometers in orbit around the Sun,
which bounce lasers between each other with displacement noise of about 10 $\mathrm{pm} / \sqrt{\mathrm{Hz}}$
in a one-way measurement. The constellation follows the Earth by about $20^{\circ}$ (Fig.~\ref{fig:configuration}).
It is expected to launch in 2030-2035, with a mission lifetime of 4 years extendable to 10 years.
Recently, some technologies have been successfully tested in the LISA pathfinder mission~\cite{Armano:2017oco}.

Taiji is a gravitational-wave space facility proposed by the Chinese Academy of Sciences~\cite{Hu:2017mde}.
The University of the Chinese Academy of Sciences and other institutes of Chinese Academy of Sciences
are involved in building it.
Like LISA, Taiji is composed of a triangle of three spacecraft with a separation distance of 3 million kilometers
in a heliocentric orbit ahead of the Earth by about $20^{\circ}$ (Fig.~\ref{fig:configuration}).
The telescope diameter will be 40 cm, the displacement noise is expected to be 8 $\mathrm{pm} / \sqrt{\mathrm{Hz}}$,
and the acceleration noise is expected to be 3 $\mathrm{fm} / \mathrm{s}^2 / \sqrt{\mathrm{Hz}}$ at 1 mHz.
Since the spacecraft are farther apart than in LISA,
Taiji is slightly more sensitive to low-frequency gravitational waves~\cite{Guo:2018npi}.
The Taiji project consists of the following three steps.
The first step is to launch a satellite to demonstrate the feasibility of the Taiji technology roadmap.
The Taiji pathfinder, officially called Taiji-1, successfully launched on 20 September 2019 and it is currently operational.
Following Taiji-1, two satellites will be launched to verify the key technologies
including long baseline interferometry in space by 2024.
Finally Taiji is planned to launch in the same period as LISA.
If Taiji joins the LISA constellation, assuming a one-year overlap,
the LISA-Taiji network in space (Fig.~\ref{fig:configuration}) is expected to significantly improve
the sky localization of gravitational-wave sources (luminosity distance and solid angle)
due to the large separation of the two constellations.

Fast and accurately localizing gravitational-wave sources is one of the key tasks
for the ground-based and space-based gravitational-wave observations.
Accurate localization is crucial for the follow-up electromagnetic spectroscopic observations
and the unique identification of their host galaxies.
With an accurate knowledge of the redshift of the host galaxy,
gravitational-wave sources can be used as standard sirens
to independently explore the expansion history of the Universe~\cite{Schutz:1986gp}.

It is hard to determine the sky location of the gravitational-wave source
using a single ground-based gravitational-wave detector
because detectors are sensitive to gravitational waves from nearly all directions.
With two detectors at different locations,
the position of the source can in principle be restricted to an annulus in the sky by triangulation
using the time difference on arrival at the two detectors.
A network of more than two detectors can localize the sky position of the source
using the arrival time difference with the help of the phase difference and amplitude ratios
of gravitational waves on arrival at the detectors.
For example, the sky localization of GW170814 was significantly improved due to the joining of the Advanced Virgo detector,
reducing the area of the 90\% credible region from 1160 deg$^2$ using only the two Advanced LIGO detectors
to 60 deg$^2$ using the LIGO-Virgo network~\cite{Abbott:2017oio}.
The addition of KAGRA and LIGO-India will further improve localization for the frequency bands covered by ground-based detectors.

Unlike ground-based detectors, the main targets for LISA and Taiji will be gravitational waves
from coalescing massive black hole binaries~\cite{Begelman:1980vb} with total masses
between $10^4$ M$_{\odot}$ and $10^8$ M$_{\odot}$ at the centres of galaxies.
In addition, LISA and Taiji might be able to detect binary stellar black holes and black hole-neutron star systems in their initial inspiral phase.
While bound massive black hole binaries are difficult to identify through electromagnetic observations,
during their inspiral phase when the orbital period of the system becomes smaller than hours,
these systems should emerge as strong gravitational-wave sources.
The inspiral would be followed by the merger of the two massive black holes,
which should be detectable by space-based gravitational-wave detectors with a high signal-to-noise ratio.
Joint electromagnetic and gravitational-wave detections of such systems will allow the study of the accretion disk
during and after the massive black hole binaries merge to a single black hole~\cite{Milosavljevic:2004cg}.

\begin{figure}[!t]
  \centering
  \includegraphics[width=5in]{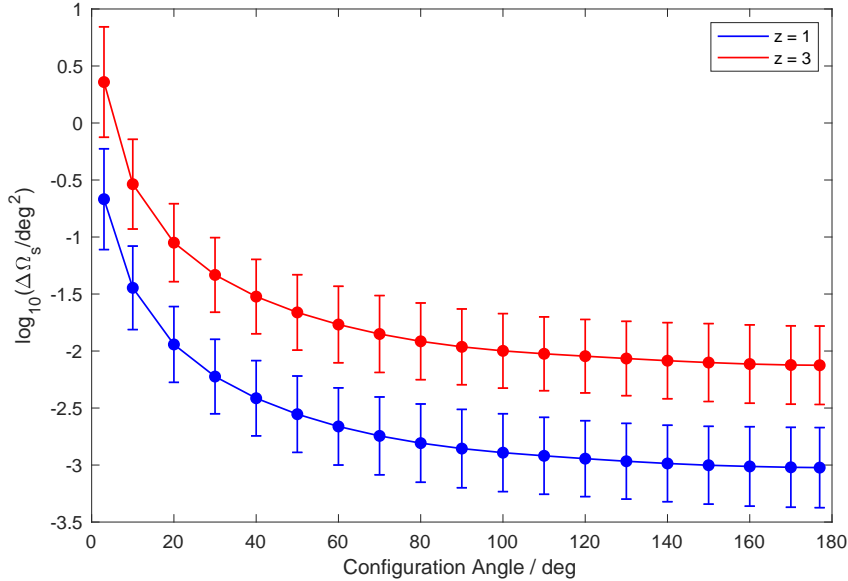}
  \caption{Measurements of the angular resolution.
  The angular resolution ($\Delta\Omega_s$) depends on the configuration angle,
  subtended by the heliocentric orbit between two detectors in the LISA-Taiji network.
  We choose an equal-mass black hole binary with a total intrinsic mass of $10^5$ M$_{\odot}$,
  located at redshifts of $z=1$ (blue) and $z=3$ (red), respectively.
  The $1\sigma$ uncertainties are evaluated using a catalogue of 10,000 simulated sources at different sky positions.}
  \label{fig:angle}
\end{figure}

The inspiral and merger of such massive black hole binaries can last between several days to years
in the frequency band of LISA and Taiji.
Due to the motion of the detectors in space,
the time dependence of the antenna pattern function plays a crucial role
in localizing the position of the gravitational-wave source~\cite{Lang:2007ge}.
Hence, a single space-based detector can be effectively treated as a network
including a set of detectors at different locations along the detector's trajectory in space,
which observe a given gravitational-wave event at different time.
LISA is expected to localize gravitational-wave sources to the angular resolution of $1-100$ deg$^2$~\cite{Audley:2017drz},
which depends on the mass, distance and inclination angle of gravitational-wave sources~\cite{Cutler:1997ta}.
However, given the expected redshift distribution of sources detectable by LISA,
such an angular resolution is not good enough to identify the source galaxy.
If Taiji joins LISA,
the LISA-Taiji network can significantly improve the sky localization of gravitational-wave,
following a similar strategy of triangulating the signal as for ground-based detectors.

The measurements of the angular resolution for the LISA-Taiji network depend on the configuration angle,
subtended by the heliocentric orbit between LISA and Taiji (Fig.~\ref{fig:angle}).
For a configuration angle of $180^{\circ}$, the angular resolution reaches a minimum value.
For an equal-mass black hole binary with a total intrinsic mass of $10^5$ M$_{\odot}$,
located at redshifts of $z=1$ and $z=3$, respectively,
the angular resolution is improved by about 2 orders of magnitude
as the configuration angle varies from $3^{\circ}$ to $40^{\circ}$
while it is improved by about 0.6 order of magnitude from $40^{\circ}$ to $180^{\circ}$.
Hence, the LISA-Taiji network with the configuration angle of $40^{\circ}$
can effectively help us to fast and accurately localize gravitational-wave sources.

\begin{acknowledgments}
We thank Wen Zhao, Jian-Min Wang, and Jan Zaanen for helpful comments.
This work is in part supported
by the National Natural Science Foundation of China Grants No.11690021, No.11690022, No.11851302, No.11747601, No.11435006 and No.11821505,
by the Strategic Priority Research Program of the Chinese Academy of Sciences Grant No.XDB23030100 and No.XDA15020701, and
by the Key Research Program of Frontier Sciences, CAS.
\end{acknowledgments}

\end{document}